\documentclass{emulateapj}
\usepackage{apjfonts}
\usepackage{natbib}

\def\Mo{{\rm M_\odot}}

\shorttitle{Pairing of SMBHs in Unequal-Mass Mergers}
\shortauthors{Callegari et al.}

\begin{document}

\title{Pairing of Supermassive Black Holes in Unequal-Mass Galaxy Mergers}

\author{Simone Callegari\altaffilmark{1}, Lucio Mayer\altaffilmark{1,2},
  Stelios Kazantzidis\altaffilmark{3}, Monica Colpi\altaffilmark{4}, \\
  Fabio Governato\altaffilmark{5}, Thomas Quinn\altaffilmark{5}, and
  James Wadsley\altaffilmark{6}} 
\email{callegar@physik.uzh.ch}

\altaffiltext{1}{Institute for Theoretical Physics, University of
  Z\"urich, Winterthurerstrasse 190, CH-9057 Z\"urich, Switzerland}
\altaffiltext{2}{Institut f\"ur Astronomie, ETH
  Z\"urich-H\"onggerberg, Wolfgang-Pauli-Strasse 16, CH-8093 Z\"urich,
  Switzerland} 
\altaffiltext{3}{Center for Cosmology and AstroParticle Physics; and
  Department of Physics; and Department of Astronomy, The Ohio State
  University, 191 West Woodruff Avenue, Columbus, OH 43210 USA}
\altaffiltext{4}{Universit\`a Milano Bicocca, Dipartimento di Fisica
  G. Occhialini, Piazza della Scienza 3, I-20126 Milano, Italy}  
\altaffiltext{5}{Department of Astronomy, University of Washington,
  Box 351580, Seattle, WA 98195, USA}
\altaffiltext{6}{Department of Physics and Astronomy, McMaster
  University, 1280 Main Street West, Hamilton, ON L8S 4M1, Canada}

\begin{abstract}
  We examine the pairing process of supermassive black holes (SMBHs)
  down to scales of 20-100~pc using a set of $N$-body/SPH simulations
  of binary mergers of disk galaxies with mass ratios of $1$:$4$ and
  $1$:$10$. Our numerical experiments are designed to represent merger
  events occurring at various cosmic epochs. The initial conditions of
  the encounters are consistent with the $\Lambda$CDM paradigm of
  structure formation, and the simulations include the effects of
  radiative cooling, star formation, and supernovae feedback.  We find
  that the pairing of SMBHs depends sensitively on the amount of
  baryonic mass preserved in the center of the companion galaxies
  during the last phases of the merger. In particular, due to the
  combination of gasdynamics and star formation, we find that a pair
  of SMBHs can form efficiently in $1$:$10$ minor mergers, provided
  that galaxies are relatively gas-rich (gas fractions of 30\% of the
  disk mass) and that the mergers occur at relatively high redshift
  ($z \sim 3$), when dynamical friction timescales are shorter.  Since
  $1$:$10$ mergers are most common events during the assembly of
  galaxies, and mergers are more frequent at high redshift when
  galaxies are also more gas-rich, our results have positive
  implications for future gravitational wave experiments such as the
  Laser Interferometer Space Antenna.
\end{abstract}


\section{INTRODUCTION}
\label{s:intro}

Compelling dynamical evidence indicates that supermassive black holes
(SMBHs) with masses ranging from $10^{6}$ to above $10^{9}$~$\Mo$
reside at the centers of most galactic spheroids
\citep[e.g.,][]{ferrareseford05}. The masses of SMBHs correlate with
various properties of their hosts, e.g.  luminosity or mass
\citep{magorrian98,haringrix04} and velocity dispersion
\citep{ferrarese00,gebhardt00}. In the currently favored model for
structure formation, the $\Lambda$CDM cosmology,
galaxies grow hierarchically through mergers and accretion of smaller
systems \citep[e.g.,][]{whiterees78}.  Thus, if more than one of the
merging galaxies contained a SMBH, the presence of two or more SMBHs
in their merger remnant will be almost inevitable during galaxy
assembly \citep{begelman80}.  However, it is unclear if the dynamical
processes at play are efficient in forming a close SMBH pair with
separations $\sim 10$--$100$~pc, which may subsequently shrink to a
bound binary, and eventually merge via gravitational wave
radiation. Such black hole coalescence events are expected to give
rise to gravitational wave bursts that should be detectable by the
Laser Interferometer Space Antenna (LISA) \citep{vecchio04}.

SMBH pairing has been shown to proceed quickly when both compact
objects are hosted by steep stellar cusps approaching each other from
close distances \citep{milomerritt01}, or when embedded in a
circumnuclear gaseous disk under appropriate thermodynamic conditions
\citep{mayer07}, but whether the large-scale merger can lead the SMBHs
to such a favorable configuration is still a matter of debate.
Previous studies found that, following a galaxy merger, the relative
distance of the SMBHs in the remnant is very sensitive to the
structure of the merging galaxies, and to their initial orbit
\citep{governato94}. \citet{kazantzidis05} showed that pairing is
efficient in equal-mass disk galaxy mergers with cosmologically
relevant orbits, while the presence of a dissipative component is
necessary for the pairing of SMBHs in $1$:$4$ mergers. Other recent
studies \citep[e.g.,][]{springel05,johansson08} focused on the effect
of energetic feedback from black hole accretion onto the surrounding
galaxy, but were not designed to follow the orbital evolution of
SMBHs. Substantially less effort has been devoted to examining the
fate of SMBHs in minor mergers \citep[but see][]{boylan07}, which are
much more frequent in $\Lambda$CDM cosmologies
\citep{laceycole93,fakhourima08}. Investigating the necessary
conditions for SMBH pair formation in this regime is of primary
importance for the search of gravitational waves and for SMBH
demographics and activity.

In this Letter, we report on the efficiency of the SMBH pairing process 
using a set of $N$-body/SPH simulations of disk galaxy mergers, with mass
ratios $q=0.25$ and $0.1$, constructed to represent mergers occurring at
various cosmic epochs. The choice of the initial
conditions, in particular the masses of the SMBHs, is such that the
corresponding SMBH coalescence events would
be detectable with LISA \citep{sesa05}. 

\clearpage
\section{SIMULATION SET-UP}
\label{s:sims}

The galaxy models were initialized as three-component systems
following the
methodology outlined in \citet{hernquist93}. They comprise a Hernquist
spherical stellar bulge \citep{hernquist90}, an exponential 
disk with a gas mass fraction $f_{\rm g}$, and an adiabatically
contracted dark matter halo \citep{blumenthal86} with an initial NFW profile 
\citep{nfw}. A collisionless particle representing the SMBH was added
at the center of each galaxy. 

Our reference model is a Milky-Way type galaxy, with a virial velocity
$V_{\rm vir}=145$ km/s, a disk mass fraction $M_d=0.04 M_{\rm vir}$,
and a bulge mass fraction $M_b = 0.008 M_{\rm vir}$. The mass of its
central SMBH is $M_{\rm BH}=2.7\times10^6$~$\Mo$, consistent with the
updated $M_{\rm BH} - M_{\rm bulge}$ relation \citep{haringrix04}. The
disk scale height and the bulge scale radius are $z_0=0.1R_d$ and
$a=0.2R_d$ respectively, $R_d$ being the exponential disk scale
length.  $R_d$ is determined following the model by \citet*{mmw98}
(MMW hereafter), which yields disk galaxies lying on the Tully-Fisher
relation.  Models at redshift $z=0$ were initialized with a halo
concentration parameter $c=12$ \citep{bullock01}.  We also ran mergers
with initial conditions rescaled to $z=3$ according to MMW, keeping
$V_{\rm vir}$ fixed, as expected for the progenitors of our $z=0$
models \citep{li07}.  Considering high-redshift mergers is crucial,
because the merger rate increases with look-back time, and a large
fraction of the gravitational wave signal from coalescences of SMBH
binaries is predicted to originate from this cosmic epoch at the
corresponding mass scale \citep{sesa05,volonteri03}.  Following MMW,
all masses, positions and softening lengths were rescaled by a factor
$H(z=3)/H_0$, i.e. the ratio between the Hubble constant at $z=3$ and
its present-day value for a $\Lambda$CDM ``concordance'' cosmology
($H_0=70$~km~s$^{-1}$~Mpc$^{-1}$, $\Omega_{\rm m}=0.3$,
$\Omega_\Lambda=0.7$).  The halo concentration was chosen according to
\citet{bullock01}, $c=3$. Satellite galaxies were initialized with the
same structure, with the mass in each component being scaled down by
$q$. The resulting SMBH pairs fall in the typical range of masses
whose coalescences will be detectable with LISA \citep{sesa05}.  We
choose orbital parameters for the mergers that are common for merging
halos in cosmological simulations \citep{benson05}: the baricenters of
the two galaxies were placed at a distance equal to the sum of their
virial radii and set on parabolic orbits with pericentric distances of
$20\%$ the virial radius of the most massive halo. All mergers we
considered were coplanar and prograde.

All simulations were performed with GASOLINE, a TreeSPH $N$-body code
\citep{gasoline}. We ran collisionless (``{\it dry}'', with $f_{\rm
  g}=0$) and gasdynamical (``{\it wet}'') mergers with the same gas
fraction in the primary and secondary galaxies, either $f_{\rm g}=0.1$
or $0.3$. In wet runs, atomic gas cooling was allowed; star formation
(SF) was treated according to \citet{stinson06}.  Gas particles are
eligible to form stars if their density exceeds 0.1~cm$^{-3}$ and
their temperature drops below $1.5\times10^4$~K, and the energy
deposited by a Type-II supernova on the surrounding gas is
$4\times10^{50}$~erg. With this choice of parameters our blast-wave
feedback model was shown to produce realistic galaxies in cosmological
simulations \citep{governato07}. A summary of our set of simulations
is presented in Table \ref{t:summary}.

In each galaxy (except for a very high-resolution test, see
\ref{ss:dry}), we employed $10^6$ particles for the halo, and,
initially, $2\times10^5$ star particles and $10^5$ gas particles, when
included.  The force softening was 100~pc in our reference model,
scaled down by $q^{1/3}$ in the satellites, and by $H(z)/H_0$ in
high-$z$ runs, yielding a force resolution of $\sim20$ pc in the
satellite galaxy for $q=0.1$ at $z=3$.  With such a high particle
number, the masses of star particles in the satellite is an order of
magnitude lower than $M_{\rm BH}$, ensuring that SMBH dynamics is not
affected by spurious two-body collisions. In what follows, we define
two SMBHs as a ``pair'' if their relative orbit shrinks down to a
separation equal to twice the softening. From these distances, a SMBH
binary may form in $\sim 1$ Myr \citep{mayer07}.

\begin{deluxetable}{ccccc}
\tablecaption{Summary of Simulations \label{t:summary}} 
\tablehead{ 
  \colhead{$q$} & 
  \colhead{SF}  & 
  \colhead{$f_{\rm g}$} & 
  \colhead{$z$} &
  \colhead{BH final distance\tablenotemark{a}} 
}
\startdata
0.25 & no & 0 & 0 & 2 -- 4 kpc  \\
0.25 & yes & 0.1 & 0 & 200 pc \\
0.1 & no & 0 & 3 & 1 -- 6 kpc \\
0.1 (hi-res) & no & 0 & 3 & 1 -- 5 kpc \\
0.1 & yes & 0.1 & 3 & 400 pc \\
0.1 & yes & 0.3 & 3 & 70 pc 
\enddata
\small{\tablenotetext{a}{When possible, estimates of pericenter and
    apocenter of the orbit 
    of the lighter SMBHs inside the merger remnant are given.}}
\end{deluxetable}

\section{RESULTS}
\label{s:results}


\subsection{Collisionless Mergers}
\label{ss:dry}

In collisionless runs, the satellite is not able to dissipate energy
gained through tidal shocks at pericentric passages
\citep{gnedin99b,taffoni03}. For $q=0.25$, dynamical friction on the
dark matter halo of the more massive galaxy is efficient, and the
satellite sinks down to a few $\sim 10$ kpc from the center after 3
orbits. At that point the central density of the satellite has
decreased considerably because of tidal heating \citep{kazantzidis04}.
Its innermost region is then tidally disrupted, leaving the small SMBH
at a distance of a few kiloparsecs. The dynamical friction timescale
has now greatly increased, because the mass of the small ``naked''
SMBH is orders of magnitude lower than that of the satellite's core
that once surrounded it. No pair is formed, and the smaller SMBH is
left wandering a few kiloparsecs away from the center of the remnant
(Fig. \ref{fig:orbits}). We note that estimating correctly the
dynamical friction timescale of the SMBH from the simulation is not
trivial in this regime, because the dark matter component is still
dynamically important at kiloparsec distances from the center of the
remnant. Even at high resolution, the mass of the dark matter
particles of the primary galaxy is comparable with that of the SMBH,
hence dynamical friction could be altered by discreteness
effects. However, the dynamical friction timescale needed for the
``naked'' SMBH to reach the center of the remnant can be estimated
using Chandrasekhar's formula \citep{colpi99}, and it turns out to
be longer than a Hubble time. Hence we conclude that the two SMBHs
will not form a pair.

In the $q=0.1$ case, dynamical friction is rather ineffective. The
sinking time for the satellite is longer than a Hubble time for a
$z=0$ merger, owing to the low initial mass of the satellite and to
mass loss due to tidal stripping \citep{colpi99}.  However, since
mergers are much more common at higher redshift, when orbital times
are shorter by a factor $H(z)/H_0$, we performed a $q=0.1$ merger
starting at $z=3$ (see \S\ref{s:sims}), an epoch at which these SMBH
pairs are predicted to be most typical. This is completed in $\sim
2.5$ Gyr. Similarly to the $q=0.25$ case, a wandering SMBH is left at
several kiloparsecs from the center of the primary (Fig.
\ref{fig:orbits}). In order to check that the tidal disruption of the
core was not affected by numerical heating, we ran the same merger
with a 5 times higher mass resolution in the stellar component of both
galaxies, and correspondingly higher force resolution; no significant
difference in the SMBH orbit was found
(Tab. \ref{t:summary}). Therefore, even in this case the two SMBHs do
not form a pair.

\subsection{Dissipational Mergers with Star Formation}
\label{ss:sf}
 
\begin{figure} 
\epsscale{1}
\plotone{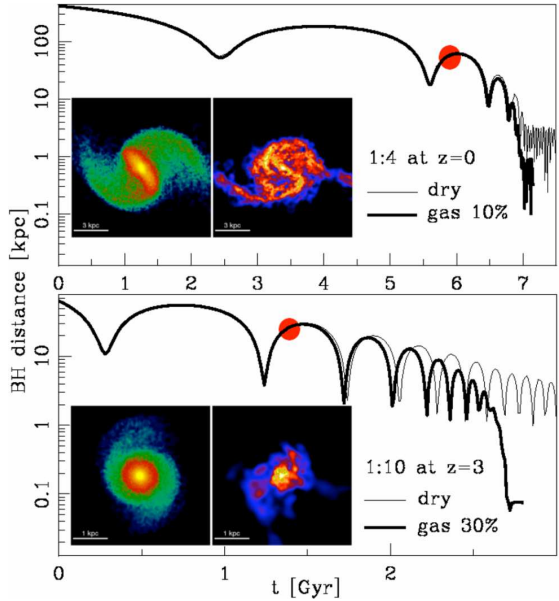}
\caption{Separation of SMBHs as a function of time in four of our
  simulations. {\it Upper panel}: SMBH distance in $q=0.25$ mergers;
  the thin and thick lines refer to the dry and wet cases, respectively. The
  inset shows the color-coded density of stars ({\it left}) and gas
  ({\it right}) for the wet case at $t=5.75$~Gyr (marked in red on the
  curve); each image is 12~kpc on a side, and colors code the range
  $10^{-2}$~--~$1$~$\Mo$~pc$^{-3}$ for stars, and
  $10^{-3}$~--~$10^{-1}$~$\Mo$~pc$^{-3}$ for the gas.
  {\it Lower panel}: SMBH distance for $q=0.1$, $z=3$; the thin and
  thick line refer to the dry and $f_g=0.3$ cases, respectively.  
  The inset shows density maps at $t=1.35$~Gyr for the $f_g=0.3$ case; 
  images are 4~kpc on a side (color coding as in upper panel).
  \label{fig:orbits}}
\end{figure}

The presence of a star-forming gaseous component crucially affects the
orbital decay of the SMBH via its dynamical response to tidal forces
and torques.

The orbits of dry and wet, $q=0.25$ mergers differ only after the
first couple of orbits ($\sim~6$~Gyr, see Fig. \ref{fig:orbits}). At
second pericenter, tidal forces excite a strong bar instability in the
satellite.  Dissipation in the gas and torques exerted by the stellar
bar onto the gas drive a gaseous inflow toward the center of the
satellite (see inset in Fig. \ref{fig:orbits}), increasing the central
star formation rate by a factor of $3$. Thus, the potential well of
the satellite deepens, ensuring resilience of its central part to
tidal stripping and shocks even when it plunges near the center of the
primary.  As a consequence, the small SMBH continues to sink fast,
because it remains embedded in the massive core of the satellite. A
pair of SMBHs is formed in this case, confirming previous results
\citep{kazantzidis05}.

\begin{figure} 
\epsscale{1}
\plotone{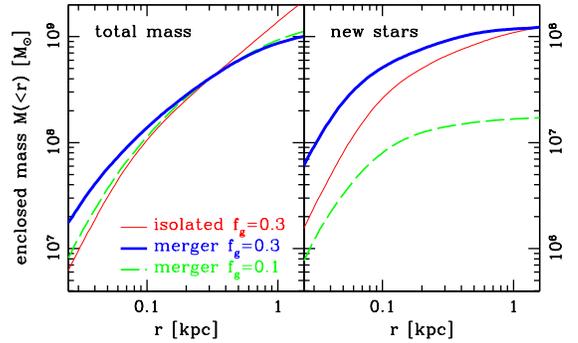}
\caption{Cumulative bound mass profiles of $z=3$, $q=0.1$
  satellites. Thin red lines refer to the $f_{\rm g}=0.3$ satellite in
  isolation, thick blue lines to the $f_{\rm g}=0.3$ merging
  satellite, and dashed green lines to the merging $f_{\rm
    g}=0.1$. All data refer to the third apocenter, or equivalent time
  in isolation ($t=1.8$~Gyr).  {\it Left panel}: total bound mass
  profiles. The more gas-rich satellite develops a higher
  concentration during the merger, compared to the other cases. Tidal
  truncation and gas removal cause a factor of $\sim2$ difference in
  mass at 2~kpc between isolation and merging cases.  {\it Right
    panel}: bound mass in stars formed after the start of the
  simulation. The total amount of stars formed depends roughly only on
  the initial $f_{\rm g}$, but SF is more localized in the center
  during the merger because of tidal forces. The more gas-rich merging
  satellite undergoes the strongest central SF burst, developing a
  higher central density.
  \label{fig:new_stars}}
\end{figure}

In $q=0.1$, $z=3$ wet mergers, both $f_{\rm g}=0.1$ and 0.3 were
employed; the latter should be a more realistic assumption, since disk
galaxies at $z=3$ are believed to have a higher gas mass fraction
\citep[e.g.,][]{franx08}. In these cases, star formation and supernovae
feedback affect the structure of the interstellar medium (ISM) in the
disks quite dramatically \citep[see also][]{governato07}. The disks
develop a clumpy and irregular multi-phase structure, and turbulent
velocities of the gas become a significant fraction (30\%) of the
circular velocity in this mass range ($V_{\rm vir} = 64$~km~s$^{-1}$).
Star formation in the center of the satellite is enhanced compared to
the same model evolved in isolation, even though the star formation
rate integrated over the whole galaxy is unchanged (see
Fig. \ref{fig:new_stars}) The concentrated star formation and the
turbulent motions of the gas stabilize the disks against bar
instability (see inset in Fig.  \ref{fig:orbits}).  In the absence of
bar-driven torques, the strong gaseous inflow seen for higher-mass
objects ($q=0.25$) does not happen for $q=0.1$. Yet, tidal torques
drive some gas towards the center, explaining the enhanced central
star formation (Fig. \ref{fig:new_stars}). During the first three
orbits, the $f_g=0.1$ case preserves its initial central density owing
to the mild mass inflow, rather than lowering it as in the
collisionless case, while the $f_{\rm g}=0.3$ satellite develops a
steeper {\it stellar} cusp. Once the satellites go through the second
pericentric passage, their ISM is prone to ram pressure stripping by
the gas disk of the primary galaxy, outside the ram pressure stripping
radius \citep{marcolini03}.  Nearly 90\% of the gas is swept away when
they first enter the disk of the primary, while what remains is
stripped during the next orbit: at $t=2$~Gyr, the satellites have lost
all their gas content, even in their central region.  From this point
onward, the satellite with initial $f_{\rm g}=0.3$ is a \emph{cuspy,
  gas-poor object}, subject to dynamical friction in the stellar and
gaseous background. Its sinking is relatively fast because the steeper
stellar density profile implies a larger bound mass, enhancing
dynamical friction relative to the dry merger case.  Moreover, its
response to tidal shocks is nearly adiabatic \citep{gnedin99b},
preserving it from tidal disruption. On the contrary, the satellite of
the $f_{\rm g}=0.1$ run undergoes a slower decay because of the lower
bound mass. It then experiences a higher number of tidal shocks at
pericentric passages which further decrease its density, until its
complete disruption. As a consequence, the $f_{\rm g}=0.3$ merger
leaves the lighter SMBH at 70~pc from the more massive one
(Fig. \ref{fig:orbits}, Tab. \ref{t:summary}) in a gas-rich
environment, where the dynamical friction timescale for the SMBH to
sink to the center, based on Chandrasekhar's formula, is very short
($<1$~Gyr).  Instead, in the $f_{\rm g}=0.1$ case the final distance
of the SMBHs is $\sim400$~pc; at such separations the dynamical
friction timescale is of a few billion years. Hence the pairing will
occur within one Hubble time in both cases, but will be considerably
faster in the $f_{\rm g}=0.3$ case.

\section{DISCUSSION AND CONCLUSIONS}
\label{s:conclusions}

Our results show that the formation of a SMBH pair in unequal-mass
mergers of disk galaxies is very sensitive to the details of the
physical processes involved. None of the collisionless cases we
studied led to SMBH pairing: tidal shocks progressively lower the
density in the satellite until it dissolves, leaving a wandering SMBH
in the remnant. The inclusion of gas dynamics and SF changes
significantly the outcome of the merger. For higher mass ratios
($q=0.25$) at $z=0$, bar instabilities funnel gas to the center of the
satellite, steepening its potential well and allowing its survival to
tidal disruption down to the center of the primary.  Therefore, in
this case the presence of a dissipative component is necessary and
sufficient to pair the SMBHs at $\sim200$~pc scales, creating
favorable conditions for the formation of a binary.  The smaller
satellites here considered ($q=0.1$, $z=3$) are more strongly affected
by both internal SF and the gasdynamical interaction between their ISM
and that of the primary galaxy. Torques in the early stages of the
merger are funnelling the gas to the center less efficiently, due to
the absence of a stellar bar and the stabilizing effect of
turbulence. As a result, ram pressure strips away all of the ISM of
the satellite. If satellites develop a higher central stellar density
by rapidly converting their gas into stars before ram pressure removes
it, they can retain enough bound mass to ensure the pairing of the two
SMBHs. Gas-rich satellites ($f_{\rm g}=0.3$) undergo a stronger burst
of SF during the first orbits, and therefore meet this requirement
better than $f_{\rm g}=0.1$ satellites. Yet in both cases the central
density of the cusp remains high enough to permit its survival,
allowing the pairing of the two SMBHs within a Hubble time.

The pairing of the two SMBHs takes less than $1$ Gyr in the gas rich
systems that should be common at $z=3$. Therefore, if the $M_{\rm
  BH}-M_{\rm bulge}$ relation approximately holds at $z=3$ as in the
local Universe, the galaxies here considered should lead to the
formation of representative SMBH pairs at such cosmic epochs
\citep{volonteri03}.  These pairs are also expected to contribute
significantly to the high-$z$ gravitational wave signal in the LISA
band \citep{sesa05}. Since we show that gas-dynamical processes allow
such an efficient pairing of the SMBHs, the results of this {\it
  Letter} strengthens the case for the observability of these
coalescence events. On the other hand, we show that the timing between
galaxy mergers and mergers of their SMBHs is sensitive to the gaseous
content of the merging galaxies. Hence, SMBH coalescence events do not
necessarily trace galaxy mergers directly.  This will have important
implications on the interpretation of the LISA data stream.

We note that the orbital evolution of the SMBH pairs, in the dynamical
range considered here, has only a weak dependence on the masses of the
two SMBHs. As shown in Figure \ref{fig:new_stars}, the stellar mass
enclosed inside two softening lengths from the center of our galaxy
models (hence close to our resolution limit) already exceeds $M_{\rm
  BH}$ by more than an order of magnitude.  This is the effective mass
that determines how quickly the SMBHs will sink. Therefore, lowering
$M_{\rm BH}$ or increasing it by up to an order of magnitude would
have no effect on sinking timescales {\it before} the disruption of
the satellite.  Instead, {\it after} disruption, the analytic estimate
for the dynamical friction timescale of the naked SMBH would change
linearly with $M_{\rm BH}$. Similarly, if nuclear star clusters with
masses $\sim10~M_{\rm BH}$ \citep{wehner06,ferrarese06} were present
around the SMBHs, their sinking timescales would still be longer than
a Hubble time in our dry mergers, where the final SMBH separation
exceeds 1~kpc, while the pairing would now occur in well below a Gyr
in \emph{all} our wet mergers. Hence, either a larger $M_{\rm BH}$ or
the presence of a nuclear star cluster would enhance even further the
difference between dry and wet mergers.

Lastly, a general limitation of our simulations is that they lack gas
accretion onto the SMBHs and associated energy feedback.  Additional
heating from the active SMBH should reduce the binding energy of the
gas, making it more susceptible to stripping processes, and perhaps
inhibiting the formation of a steep stellar cusp. This would reduce
the efficiency of the pairing process, but the effect will strongly
depend on when the SMBH becomes active during the merger. Although it
is unlikely that these effects will change the overall picture
presented in this Letter, they will have to be explored in a
forthcoming paper.

\acknowledgments The authors are grateful to B. Devecchi, M. Dotti,
A. Maller, D. Merritt, D. Weinberg, and S. White for fruitful
discussions, and to D. Potter for technical support.  The simulations
were performed on the zBox2 and zBox3 supercomputers at the ITP,
University of Z\"urich. SK is supported by the CCAPP at The Ohio State
University. MC thanks the Aspen Center for Physics for its hospitality
while this work was in progress.  FG is supported by the NSF grant
AST-0607818. TQ and FG are supported by NASA grant NNX07AH03G.

\newpage

\bibliography{}

\end{document}